# Leveraging vehicle connectivity and autonomy to stabilize flow in mixed traffic conditions: accounting for human-driven vehicle driver behavioral heterogeneity and perception-reaction time delay

Yujie Li[a,#], Sikai Chen [b,#,*], Paul (Young Joun) Ha[c], Jiqian Dong[d], Aaron Steinfeld[e], Samuel Labi[f]

*Abstract*—The erratic nature of human driving tends to trigger undesired waves that amplify as successive driver reactions propagate from the errant vehicle to vehicles upstream. Known as phantom jams, this phenomenon has been identified in the literature as one of the main causes of traffic congestion. This paper is based on the premise that vehicle automation and connectivity can help mitigate such jams. In the paper, we design a controller for use in a connected and autonomous vehicle (CAV) to stabilize the flow of human-driven vehicles (HDVs) that are upstream of the CAV, and consequently to lower collision risk in the upstream traffic environment. In modeling the HDV dynamics in the mixed traffic stream, we duly consider HDV driver heterogeneity and the time delays associated with their perception reaction time. We can find that the maximum number of HDVs that a CAV can stabilize is lower when human drivers potential time delay and heterogeneity are considered, compared to the scenario where such are not considered. This result suggests that heterogeneity and time delay in HDV behavior impairs the CAVs capability to stabilize traffic. Therefore, in designing CAV controllers for traffic stabilization, it is essential to consider such uncertainty-related conditions. In our demonstration, we also show that the designed controller can significantly improve both the stability of the mixed traffic stream and the safety of both CAVs and HDVs in the stream. The results are useful for real-time calibration of the model parameters that characterize HDV movements in the mixed stream – this knowledge is essential for effective real-world deployment of CAV controllers in mixed traffic stream environments.

## 1. INTRODUCTION

In the United States, as in several other countries, traffic congestion continues to pose a serious problem, with profound adverse effects on energy consumption and emissions. The transportation sector accounted for 28% of total U.S. energy consumption in 2018; of this, more than 75% is attributed to the highway mode. On average, the U.S. commuter wasted nearly 7 full working days in 2017 sitting in congested traffic, which translates to over $1,000 in personal costs (Schrank et al., 2019). Traffic delay translates into reduced economy productivity of transport-dependent and transport-related industries, and driver frustration and fatigue. For these reasons, highway agencies at all levels of government continue to seek opportunities to improve the efficiency of their transportation system from congestion-related perspectives including, energy consumption, emissions, and productivity.

In efforts to mitigate congestion and therefore reduce its accompanying economic and social costs, extensive studies have been carried out by seeking to identify the underlying causes of congestion. In this regard, some studies have classified the causes of congestion, at least from the supply side, as follows: bottleneck related triggers and non-bottleneck related triggers. Bottleneck-related triggers are associated with capacity reduction due to decreased number of lanes, and merging or grade change locations (Laval, 2006; Laval and Daganzo, 2006; Zheng et al., 2011). Non-bottleneck related triggers are associated with congestion that arises in the absence of physical bottlenecks and include *phantom traffic jams* (Orosz et al., 2009). Phantom jams occur because human drivers sometimes demonstrate irrational or spontaneous driving patterns. These are caused by intended or unintended driving or non-driving related actions or inactions during the driving task. Examples of such driving behavior could include sudden acceleration or deceleration, unexpected braking or lane changing, or hesitating (Horn and Wang, 2018). When a driver of a vehicle engages in such errant behavior, responsive actions are sparked spontaneously in the chain of vehicles behind it, as they seek to avoid a crash, particularly in dense traffic where headways are small. The resulting perturbations become amplified to form stop-and-go waves that travel backwards along the platoon. In traffic flow, this property is known as *string instability* (Burnham et al., 1974; Sugiyamal et al., 2008). A number of experimental studies have reproduced the degradation in traffic string stability that leads to such phantom jams (Jiang et al., 2018, 2014; Stern et al., 2018; Sugiyamal et al., 2008; Tadaki et al., 2013) but did not provide explicit solutions to avoid these jams. Fortunately, with the advent of connected and autonomous vehicles (CAVs), there seems to be a promising solution to this traffic operations challenge. With vehicle automation and connectivity-aided communication, the vehicle is afforded enhanced awareness of its surrounding traffic conditions. It has been shown that CAVs can help reduce congestion, increase safety, improve productivity, and increase the capacity of existing transportation facilities (Talebpour and Mahmassani, 2016). This is consistent with the anticipation that CAVs can help resolve some longstanding transportation engineering problems as stated in recent publications by USDOE (US Energy Information Administration, 2017) and AASHTO, ITE, and ITSA (AASHTO/ITE/ITSA, 2020).

# The first two authors contributed equally to the study.
a. Center for Connected and Automated Transportation (CCAT), and Lyles School of Civil Engineering, Purdue University, W. Lafayette, IN 47907.
b*. Center for Connected and Automated Transportation (CCAT), Lyles School of Civil Engineering, Purdue University, W. Lafayette, IN 47906, and Robotics Institute, School of Computer Science, Carnegie Mellon University, Pittsburgh, PA (corresponding author, phone: 213-806-0141; e-mail: chen1670@purdue.edu; sikaic@andrew.cmu.edu

c. Center for Connected and Automated Transportation (CCAT), and Lyles School of Civil Engineering, Purdue University, W. Lafayette, IN 47907.
d. Center for Connected and Automated Transportation (CCAT), and Lyles School of Civil Engineering, Purdue University, W. Lafayette, IN 47907.
e. Robotics Institute, School of Computer Science, Carnegie Mellon University, Pittsburgh, PA 15213.
f. Center for Connected and Automated Transportation (CCAT), and Lyles School of Civil Engineering, Purdue University, W. Lafayette, IN 47907.

In the literature, most studies that evaluated the safety and mobility benefits of CAVs are predicated on the assumption of full CAV market penetration (Gunter et al., 2019; Milanes et al., 2014; Schakel et al., 2010). While the findings of these studies represent pioneering efforts and are innovative, the assumption of CAV full market penetration may be rather unduly restrictive. This is because, in reality, the expectation is that CAVs will be ushered into the market in an incremental fashion, and full market penetration is expected to occur only in the distant future after a lengthy transition (TRB, 2017; Litman, 2019). In other words, it is anticipated that after their market entry, CAVs will coexist with human-driven vehicles (HDVs) to form a "mixed" traffic stream, for a long time. Unlike CAVs whose movements can be well planned and designed in advance, and therefore are relatively foreseeable, HDV movements involve a higher level of unpredictability and heterogeneity that can introduce uncertainties and disturbances into the mixed traffic flow as explained in an earlier paragraph. A number of studies have argued that in a mixed traffic stream where CAVs share the road with HDVs, the latter will impair the performance of not only CAV flows but also the overall traffic stream in its entirety (Milanés et al., 2013; van den Broek et al., 2011). Additionally, some studies examined impacts of CAV in mixed traffic flow conditions with simulated environment (Shladover et al., 2001, 2012; VanderWerf et al., 2002). These works are intuitive, but in most papers, the interactions between CAVs and HDVs are assumed to be unchanged. For this reason, it is essential to design a CAV controller that addresses not only the mixed traffic situation but also the uncertainties associated with HDVs. Such a controller, by mitigating the effects of the uncertainties and disturbances in mixed traffic flow, can improve the traffic stream performance.

In the literature, researchers have determined that the traffic flow of the entire string can be stabilized by controlling the behavior of certain vehicles in the platoon (Wu et al., 2018; Stern et al., 2018; Chen et al., 2020). More specifically, Wu (2018)'s pioneering study stated that for autonomous vehicles specifically introduced into the traffic stream (or already extant in the traffic stream) to address traffic string instability, a threshold of 6% uniform market penetration can effectively dissipate the oscillations and thereby stabilize the traffic under all uniform flow conditions. The Wu study was based on two key assumptions: (1) HDV dynamics are homogeneous, and (2) time delay between the vehicles actions in the string are negligible.

In an effort to build on the previous research in this area, the present paper identified a number of research opportunities and seeks to make research contributions in these respects:

- **Considering mixed (rather than uniform) traffic flow**
  Researchers have claimed that CAVs can potentially enhance the efficiency of traffic operations and have demonstrated this experimentally in a fully-HDV or fully-CAV environment. At least one researcher went further: Wu et al. (2018) explored the modeling approaches that characterize traffic flow dynamics and the impacts on stability in a mixed traffic stream of CAVs and HDVs. It is important to pursue research directions consistent with Wu et al's work because that reflects a more realistic prognostication of the state of traffic in the expectantly lengthy transition period of CAVs.

- **Accounting for the behavioral heterogeneity of HDV drivers**
  Previous studies that examined autonomous system control logic for mixed traffic conditions, had assumed homogeneous behavior of HDVs. However, this is a rather restrictive assumption because in reality, human driver behavior is inherently heterogeneous as explained in a preceding section of this paper. For example, given different levels of aggression and varied reaction times, the actions taken by drivers will differ greatly.

- **Incorporating the perception-reaction time delay associated with the human driver (HDV operations)**
  Another strong assumption made in most past studies on system control logic for mixed traffic conditions is that in modeling HDV dynamics, there is no time delay in the information perception and driver reaction. This assumption helps to simplify the problem settings but seems rather unrealistic because humans' perception-reaction times are often significant. By nature, humans are unable to react to stimuli in the exact instance they receive it. Also, there is an inherent delay in HDV braking reactions. For HDVs, the trailing driver must first perceive the tail-light illuminating, track the change in speed, decide to brake, and then activate/move their foot. This delay leads to a need to brake harder to preserve the time gap due to the latency in human perception and decision making (Krishnan et al., 2001). Considering such delay is important because some studies found that stable platoon size decreases when greater time delay is introduced into the system (Bando et al., 1998). Therefore, the practical effectiveness of CAV controllers developed in past studies to stabilize traffic, may be limited.

In the present paper, we address these limitations of past studies in this area by considering a more realistic traffic environment. First, we propose a design for CAV control that stabilizes traffic flow and improves overall mobility in a mixed traffic stream. Secondly, to model the HDVs in the traffic stream more realistically, we duly consider the HDV drivers perception-reaction time and the heterogeneity of their driving behavior. For example, the car-following behavior of an HDV may be very different when following different classes of vehicles such as a small automobile versus a large truck. Listing all the possible combinations of vehicle class and follower-followed order in the platoon may seem necessary (Krishnan et al., 2001). However, not only is such enumeration time-consuming and tedious but also may not be necessary in the proposed framework. This is because in the framework, the HDV dynamics are calibrated in real time due to connectivity. The control system described in paper's framework can be calibrated and updated in real time, which are critical requirements for safe and efficient CAV operations.

The rest of the paper is organized as follows. Section 2 reviews related work in literature. The mathematical models for HDV and CAV and the assumptions and consideration of stability and safety are presented in Section 3. Based on that, the controller design problem is formulated into optimization problem in Section 4. We validate the proposed controller using a numerical experiment in Section 5. Section 6 discusses the results and summarizes the study's findings.

## 2. RELATED WORK

### 2.1 Autonomous vehicle control

Adaptive cruise control (ACC) is a concept that involves automatic adjustment of a vehicle's cruise-control velocity (in the presence of downstream traffic) to a safe following distance. As a partially automated driving feature, ACC seeks to enable longitudinal control of the vehicle and to reduce the driver's workload (Vahidi and Eskandarian, 2003; Ioannou and Chien, 1993). Over the years, rapid developments in information and communication technologies have yielded promising extensions of the ACC concept to a cooperative system feature known as CACC (cooperative adaptive cruise control). CACC systems leverage the availability of vehicle-to-vehicle (V2V) communications to collect more extensive and reliable information (Milanés et al., 2013). Doing this promotes enhanced awareness of the surrounding traffic environment and thereby improves the control system's reliability and performance. Some previous researchers have developed conceptual CACC models to evaluate their feasibility in achieving traffic safety and efficiency (Kato et al., 2002).

The ACC controller generally uses an error-based controller that attempts to ensure that the levels of two parameters are maintained in an equilibrium state: $v_i = v_{i-1}$ and $\Delta x_i = \Delta x_i^*$, where $v_i$ is the velocity of vehicle $i$, $\Delta x_i$ is the car-following headway of vehicle $i$ and $\Delta x_i^*$ denotes the desired headway. The ACC control model (equation (1)), represents the driver's desire to maintain a desired headway between the preceding vehicles that have the same velocity. Given the gaps from the desired headway and deviation from the desired speed, the driver responds to this situation and to reduce the deviation by either acceleration or deceleration through the following model (Equations 1-3):

$$a_i(t) = K_v\big(v_i(t) - v_{i-1}(t)\big) + K_p\big(\Delta x_i(t) - \Delta x_i^*(t)\big) \\ = K_v\big(\dot{x}_i(t) - \dot{x}_{i-1}(t)\big) + K_p\big(\Delta x_i(t) - \Delta x_i^*(t)\big) \quad (1)$$

s.t.,

$$a_{min} \leq a_i(t) \leq a_{max} \quad (2)$$
$$\Delta x_i^*(t) = \lambda_1\big(v_i^2(t) - v_{i-1}^2(t)\big) + \lambda_2 v_i(t) + \lambda_3 \quad (3)$$

where: $a_{max}$ and $|a_{min}|$ denotes the maximum acceleration and deceleration rates, respectively; $K_v$, $K_p$ are positive controller gains that are to be tuned. $\lambda_1, \lambda_2, \lambda_3$ are the constants for the safety design policy. Equation (2) specifies constraints on acceleration considering physical limitations and passenger comfort. Equation (3) defines the desired headway. For tight vehicle-following situations, equation (3) can be re-written as: $\Delta x_i^* = \lambda_2 v_i + \lambda_3$, which combines the constant-space headway policy ($\Delta x_i^* = \lambda_3$) or the constant-time headway policy ($\Delta x_i^* = \lambda_2 v_i$).

The CACC controller, similar to the ACC system, seeks to maintain three invariants: $v_i = v_{i-1}$, $\Delta x_i = \Delta x_i^*$ and $a_i = a_{i-1}$. The CACC control model can be expressed by equation (4):

$$a_i(t) = K_v\big(\dot{x}_i(t) - \dot{x}_{i-1}(t)\big) \\ + K_p\big(\Delta x_i(t) - \Delta x_i^*(t)\big) + K_a a_{i-1} \quad (4)$$

Similar to ACC model (equation (1)), the first two terms in the right-hand side of Equation (4) consider the velocity difference and deviation from desired headways, and the third term distinguishes the CACC controller from the ACC using additional acceleration information from the direct predecessor. (Wilmink et al., 2007) extended this CACC system by considering multi-predecessors (Equation 5):

$$a_i(t) = K_v\big(\dot{x}_i(t) - \dot{x}_{i-1}(t)\big) + K_p\big(\Delta x_i(t) - \Delta x_i^*(t)\big) \\ + \frac{K_v}{n-1}\sum_{k=i-n+1}^{i-1}\big(\dot{x}_k(t) - \dot{x}_{k-1}(t)\big) \quad (5)$$

### 2.2 Traffic dynamics

The driving behavior of conventional (or, human-driven) vehicles has been studied extensively since the early 50's (Bekey et al., 1977; Burnham et al., 1974; Chandler et al., 1958; Pipes, 1953; Tyler, 1964). In car-following behaviors, the human driver's actions can be likened to a controller that senses velocities, gaps, and accelerations of the vehicles in its vicinity, and accordingly makes control output decisions. The Optimal Velocity Model (OVM) (Bando et al., 1998, 1994; Nakayama et al., 2002) and the Intelligent Driver Model (IDM) are widely used models for simulating HDV traffic dynamics using CACC systems.

### C. Metrics for controller performance

#### 1) String stability:

- Vehicular string stability: Generally, the string stability of the system implies "uniform boundedness of the states" of all vehicles in the platoon (Swaroop, 1996). Typical manifestations of string instability include the transients caused by a lead vehicle's speed changes which may be amplified, leading to "slinky-type effects" upstream (Hedrick et al., 1991; Sheikholeslam and Desoer, 1992).
- Platoon string stability: In car-following dynamics, each vehicle, together with others in its vicinity, is considered a "coupled system". The stability of an individual vehicle does not necessarily translate into the stability of the entire stream of vehicles. Therefore, the concept of platoon string stability is used to define the boundaries of the propagated perturbation.

This paper considers two levels of stability: vehicle-level stability and system-level string stability. A key prerequisite in addressing the phantom traffic jam problem is to assess whether a specified platoon of vehicles is string stable. In a bid to do this, previous researchers assumed that the platoon is homogeneous. This helps to simplify the problem to a large extent, and makes it possible to make prescriptions for string stability by analyzing the dynamics of a single pair of vehicles only (Gunter et al., 2019). Unfortunately, such an assumption is not consistent with the realities of a mixed traffic stream and a stream where the heterogeneity of the human driving behaviors is significant.

#### 2) Safety measurements:

In our framework, it is essential to measure the level of safety associated with controller performance. The number of crashes is a common metric for doing this. However, due to the preponderance of traffic conflicts compared to crashes, the latter is not preferred. The application of Traffic Conflict Techniques (TCTs) to traffic safety analysis has seen considerable research interest and has been viewed as a proactive surrogate approach (Tarko, 2018; Zheng et al., 2014). This concept provides a dimension of crash severity along which all traffic events, conflict or non-conflict, can be accommodated. There are two main classes of proximity measures: temporal proximity and spatial proximity. Temporal proximity indicators, which are the most prominent and widely-used indicators, include the time-to-collision (TTC), time exposed time-to-collision (TET), time integrated time-to-collision (TIT) (Hydén, 1996; Minderhoud and Bovy, 2001).

## 3. METHODOLOGY

### 3.1 Preliminaries

The standard schema of a platoon (Fig. 1) comprises a leading vehicle (vehicle 0) and mixed traffic (both connected human-driven vehicles (HDVs) and connected and autonomous vehicles (CAVs)) that follow (or, are upstream of) the leading vehicle. The leading vehicle has unpredictable movements that is the source of the string instability.

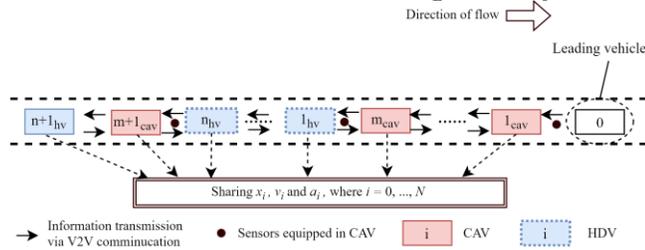

Fig. 1. Standard platoon schema in a mixed traffic stream

To ensure a rigorous inquiry into the issue, we make the following assumptions for a platoon: 1) All the vehicles in the traffic stream (HDVs and AVs) are well-connected; 2) the CAVs have no communication error or propagation time delay. The first assumption can be ensured when defining platoons that the total $N$ vehicles in the platoon should be within communication range. The second assumption can be considered realistic because there exist promising innovative technologies, such as 5G, that facilitate such efficiency in vehicle-to-vehicle communications. Therefore, it is expected that the issue of communication error and delay can be addressed.

### 3.2 Dynamics of the human-driven vehicles

To model the dynamics of traffic flow, microscopic models are used to describe the motion of individual vehicle after a stimulus, such as acceleration or deceleration, is applied to the vehicle. The dynamics of a standard car-following phenomenon can be expressed as (Equation 6):

$$\ddot{x}_i = f(\dot{x}_i, \Delta x_i, \Delta \dot{x}_i) \tag{6}$$

where:

$x_i, \dot{x}_i$ and $\ddot{x}_i$ denote the displacement, velocity and acceleration of the $i$-th vehicle. Based on (6), the acceleration of vehicle $i$ therefore depends on its velocity $\dot{x}_i$, the headway $\Delta x_i := x_{i-1} - x_i$ and the velocity difference, $\Delta \dot{x}_i$, between the preceding vehicle.

In particular, the optimal velocity model is used to model the car following dynamics of conventional vehicles (that is, HDVs). The model can be expressed as (Equation 7):

$$\ddot{x}_i = \alpha[V(\Delta x_i) - \dot{x}_i] \tag{7}$$

where: $\alpha$ represents the driver's sensitivity. $V(\Delta x_i)$ denotes an optimal velocity, which is a function of the headways. The model describes the following behaviors: the driver perceives the gaps and determines an optimal velocity at which the driver desires to travel. However, in most of the cases, there exists a deviation between the optimal and the current velocity. Awareness of such deviation stimulates the driver to reduce the deviation by accelerating or decelerating.

However, the original Optimal Velocity (OV) model does not explicitly account for driver response time which could introduce time delay in the HDV's dynamics. Bando et al. (1998)'s modification of the OV model explicitly considers delay, and the system's state-space description (Equation 7) can be written as (Equation 8):

$$\ddot{x}_i(t) = \alpha_i \left( V(\Delta x_i(t - \tau_i)) - \dot{x}_i(t - \tau_i) \right) \tag{8}$$

As shown in Equation (8), due to the natural reaction time of human drivers, the HDV will react to stimuli not at the exact time it receives the stimuli but after several seconds. Additionally, due to heterogeneity in reaction times across the HDV driver population, the sensitivity parameter $a_i$ and the time delay parameter $\tau_i$ are not necessarily the same for different HDVs. Helbing and Tilch (1998) examined this issue. Fitting the model with empirical data, they found that this model may lead to some unrealistic behavior including sharp acceleration and deceleration actions. To overcome this limitation, (Jiang et al., 2001) proposed a full velocity difference model (FVDM), as follows (Equation 9):

$$\ddot{x}_i(t) = \alpha_i \left( V(\Delta x_i(t - \tau_i)) - \dot{x}_i(t - \tau_i) \right) + \beta_i \Delta v(t - \tau_i) \tag{9}$$

The parameters $\alpha_i$ and $\beta_i$ reflect the relative weights associated with the state of vehicle $i$ in traveling at the optimal velocity and the state of the vehicle in following the preceding vehicle. The stability of the non-linear system is then analyzed by considering the equilibrium state. When the flow is uniform, all the HDVs travel at the same velocity $v^*$ with desired headways $\Delta x_i^*$. The desired headways represent an encapsulation of the constant time headway rule and the constant clearance rule. A Taylor expansion of the full velocity difference model in the vicinity of the equilibrium point, yields Equation (10):

$$\ddot{x}_i(t) = \alpha_i V'(\Delta x_i^*(t))(\Delta x_i(t - \tau_i) - \Delta x_i^*) - \alpha_i(\dot{x}_i(t - \tau_i) - v^*) + \beta_i \Delta \dot{x}_i(t - \tau_i) \tag{10}$$

It is required that the state variables are bounded. Therefore, the modified variables $\tilde{x}_i(t) := x_i(t) - x_i^*(t)$ were used. Then, Equation (10) becomes (Equations 11-13):

$$\Delta x_i^*(t) := \lambda_{2i} v_i(t) + \lambda_{3i} \tag{11}$$

$$x_i(0) := -\sum \Delta x_i^*(0) = -\sum (\lambda_{2i} v^* + \lambda_{3i}) \quad \forall i \tag{12}$$

$$x_i^*(t) := x_i(0) + t v^* \quad \forall i \tag{13}$$

The linearized system can be re-written as (Equation 14):

$$\ddot{\tilde{x}}_i(t) = k_{1i}\left(\tilde{x}_{i-1}(t-\tau_i) - \tilde{x}_i(t-\tau_i) - \lambda_{2i}\dot{\tilde{x}}_i(t-\tau_i)\right) - k_{2i}\dot{\tilde{x}}_i(t-\tau_i) + k_{3i}\left(\dot{\tilde{x}}_{i-1}(t-\tau_i) - \dot{\tilde{x}}_i(t-\tau_i)\right) \quad (14)$$

where:

$k_{1i} = \alpha_i V'(\Delta x_i^*)$, $k_{2i} = \alpha_i$, and $k_{3i} = \beta_i$

After conducting a Laplace transformation of $\tilde{x}_i(t)$ and denoting the states as $\tilde{X}_i(s)$, the transfer function for (14), $T_i(s)$ can be derived as follows (Equations 15 and 16):

$$s^2 \tilde{X}_i(s) = k_{1i}\left(e^{-s\tau_i}\tilde{X}_{i-1}(s) - e^{-s\tau_i}\tilde{X}_i(s) - \lambda_{2i}se^{-s\tau_i}\tilde{X}_i(s)\right) - k_{2i}se^{-s\tau_i}\tilde{X}_i(s) + k_{3i}se^{s\tau_i}\left(\tilde{X}_{i-1}(s) - \tilde{X}_i(s)\right) \quad (15)$$

$$T_i(s) = \frac{\tilde{X}_i(s)}{\tilde{X}_{i-1}(s)} = \frac{(k_{1i} + sk_{3i})e^{-s\tau_i}}{s^2 + s(k_{2i} + k_{3i} + k_{1i}\lambda_{2i})e^{-s\tau_i} + k_{1i}e^{-s\tau_i}} \quad (16)$$

### 3.3 Dynamics of CAVs

For purposes of modeling the CAV dynamics, we use the full velocity difference model without-time-delay (Jiang et al., 2001) (Equation 17):

$$\ddot{x}_i(t) = k_1\left(\Delta x_i(t) - \Delta x_i^*(t)\right) + k_2(\dot{x}_i(t) - v^*) + k_3\Delta \dot{x}_i \quad (17)$$

Similar to the mathematical manipulation made to the HDV model, the variable $\tilde{x}_i(t)$ can be transformed as follows:

$\tilde{x}_i(t) = x_i(t) - x_i^*(t)$, and then Equation (17) can be re-arranged to yield (Equation 18):

$$\ddot{\tilde{x}}_i(t) = k_1\left(\tilde{x}_{i-1}(t) - \tilde{x}_i(t) - \lambda_2\dot{\tilde{x}}_i(t)\right) - k_2\dot{\tilde{x}}_i(t) + k_3\left(\dot{\tilde{x}}_{i-1}(t) - \dot{\tilde{x}}_i(t)\right) \quad (18)$$

Therefore, the transfer function for CAVs can be expressed as (Equation (19)):

$$T_A(s) = \frac{\tilde{X}_i(s)}{\tilde{X}_{i-1}(s)} = \frac{(k_1 + sk_3)}{s^2 + s(k_2 + k_3 + k_1\lambda_2) + k_1} \quad (19)$$

### 3.4 Controller performance metrics

#### 3.4.1 Vehicular String stability ($\mathcal{L}_2$ stability)

To analyze the string stability of the mixed traffic platoon, the frequency-domain approach is used. For each human-driven vehicle $i$, substitute $s$ with $j\omega$, where $\omega$ (frequency), is $\geq 0$. The magnitude-squared frequency response can be obtained:

$$|T_i(j\omega)|^2 = \frac{k_{1i}^2 + \omega^2 k_{3i}^2}{\omega^2 K^2 + \omega^4 + k_{1i}^2 + f(\omega)} \quad (20)$$

where $K = k_{2i} + k_{3i} + k_{1i}\lambda_{2i}$, $f(\omega) = -2\omega^3 K\sin\omega\tau_i - 2\omega^2 k_{1i}\cos\omega\tau_i$. To ensure the uniform boundedness for the HDVs, $|T(j\omega)| \leq 1$, $\forall \omega \geq 0$. Re-arranging (20), yields the following inequality (Equation 21):

$$\omega^4 - 2\omega^3 K \sin\omega\tau_i + \omega^2(K^2 - k_{3i}^2 - 2k_{1i}\cos\omega\tau_i) \geq 0 \quad (21)$$

Equation (21) is analytically intractable and therefore cannot yield a simple solution. Therefore, rather than directly find critical value for $\omega$, sufficient conditions for string stability are sought. The conditions $\omega\tau_i \geq \sin\omega\tau_i$ and $\cos\omega\tau_i \leq 1$ hold for $\forall \omega, \tau_i \geq 0$. If $\omega$ satisfies Equations 22 and 23, then the string stability for the HDVs can also be guaranteed:

$$\omega^2 - 2\omega^2 K\tau_i + K^2 - k_{3i}^2 - 2k_{1i} \geq 0 \quad (22)$$

$$(1 - 2K\tau_i)\omega^2 \geq k_{3i}^2 + 2k_{1i} - K^2 \quad (23)$$

For the special case when $\tau_i = 0$ and $\lambda_2 = 0$, Equation (23) can be re-arranged yield Equation 24:

$$\omega^2 \geq 2k_1 - k_2^2 - 2k_2k_3 \quad (24)$$

This inequality is consistent with the string stability condition for the linearized dynamics of a non-delayed system (Ioannou and Xu, 1994; Orosz et al., 2010). While when $\tau_i \neq 0$, the parameters $k_{1i}$, $k_{2i}$ and $k_{3i}$ satisfying equation (24) in the non-delayed system do not necessarily guarantee string stability. From Equation (23), if the time delay term $\tau_i$ for vehicle $i$ is large enough, i.e., $\tau_i \geq \frac{1}{2K}$, with any disturbance at frequency $\omega$, the HDV can not maintain string stability. In other words, even with a small disturbance, the perturbations cannot be dissipated and will be amplified. To ensure that HDVs are string stable given some $\omega$, $(1 - 2K\tau_i) > 0$ and $k_{3i}^2 + 2k_{1i} - K^2 \geq 0$ should be checked initially. With these two inequalities being satisfied, the critical frequency for $i$-th HDV model is given by (Equation 25):

$$\omega_{i0}^H = \sqrt{\frac{k_{3i}^2 + 2k_{1i} - K^2}{(1 - 2K\tau_i)}} \quad (25)$$

Based on the string stability analysis above, the $i$-th HDV is string unstable given a perturbation with frequency $\omega$, $0 < \omega < \omega_{i0}^H$. For the HDV platoon that consists of multiple HDVs, the most critical frequency is given by (Equation 26):

$$\omega_0^H = \min_i \omega_{i0}^H \quad (26)$$

Carrying out a similar variable substitution $s = j\omega$ to CAV model, it is determined that the magnitude-squared frequency response is given by (Equation 27):

$$|T_A(j\omega)|^2 = \frac{k_1^2 + \omega^2 k_3^2}{(k_1 - \omega^2)^2 + \omega^2(k_2 + k_3 + k_1\lambda_2)^2} \quad (27)$$

To ensure string stability of CAVs for $\forall \omega$, $k_1, k_2$ and $k_3$ should be tuned to satisfy $|T_A(j\omega)| \leq 1$. This means that (Equation 28):

$$\omega^4 + \omega^2(k_2^2 + k_1^2\lambda_2^2 + 2k_2k_3 + 2k_1k_2\lambda_2 + 2k_1k_3\lambda_2 - 2k_1) \geq 0 \quad (28)$$

Simplifying (28) yields the requirement for CAV string stability for all perturbations, which can be expressed as (Equation 29):

$$k_2^2 + k_1^2\lambda_2^2 + 2k_2k_3 + 2k_1k_2\lambda_2 + 2k_1k_3\lambda_2 - 2k_1 \geq 0 \quad (29)$$

Therefore, appropriate levels of $k_1$, $k_2$, and $k_3$ should be selected based on Equation (29), to ensure the string stability of CAVs. The condition (29) is consistent with the well-known conditions (Wilson and Ward, 2011).

*3.4.2   Platoon string stability ($\mathcal{L}_2$ weak string stability)*

We demonstrate requirements for vehicular stability, but as mentioned in previous section, the stability of an individual vehicle does not necessarily translate into the stability of the entire stream of vehicles. Thus, we further require platoon string stability, which can be expressed mathematically as Equation (30):

$$\left\|\prod_{\forall i} T_i(j\omega)\right\|_\infty \leq 1 \quad (30)$$

The platoon string stability is essential in mixed traffic streams. Specifically, with some unstable oscillations triggered by the uncontrolled leading vehicle (vehicle 0 in Fig. 1), the autonomous vehicle can dampen the shockwave and thereby eliminate the propagation of unstable waves. Assume that for any perturbation of frequency $\omega$, $\omega \in (0, \omega_0^H)$, the CAV can stabilize the $n$ HDVs that follow it. From Equation (30), Equations (31)-(33) can be derived:

$$\left|T_A(j\omega)\prod_i^n T_i(j\omega)\right| \leq 1 \quad (31)$$

$$log|T_A(j\omega)| + \sum_i^n log|T_i(j\omega)| \leq 0 \quad (32)$$

$$log|T_A(j\omega)| + \sum_i^{n+1} log|T_i(j\omega)| > 0 \quad (33)$$

The maximum number of stabilized HDVs, $n^*_{stable}$, under $\forall \omega \in (0, \omega_0^H)$, is given by:

$$n^*_{stable} = min\{\min_{\omega}\{\left(log|T_A(j\omega)| + \sum_i^n log|T_i(j\omega)|\right) \cdot \left(log|T_A(j\omega)| + \sum_i^{n+1} log|T_i(j\omega)|\right) \leq 0\}\} \quad (34)$$

*3.4.3   Safety considerations:* The achievement of string stability does not guarantee that the system is collision-free. Additionally, it does not eliminate extremely conservative and inefficient conditions (which happens when the CAV attempts to maintain a large headway). Therefore, besides using a frequency-domain approach, it is important to impose constraints on the headway (Wu et al., 2018). Then, there exist two headway-related safety considerations:

- Minimum headway $\Delta x^-$: This ensures that the vehicle is free from collision and is equal to an effective vehicle length (Bando et al., 1998). Consider the traffic condition where each driver maintains an extra distance margin that might be needed for stopping to avoid collision. Then, the effective vehicle length exceeds the actual vehicle length.
- Maximum headway $\Delta x^+$: This ensures that the vehicle will not maintain an unreasonably large headway because that will result in low throughput and therefore, traffic inefficiency.

Based on the above considerations, the constraints for the headways can be represented as:

$$\Delta x^- \leq \Delta x_i \leq \Delta x^+ \quad (35)$$

Then, for a specific disturbance with magnitude $\beta$, Wu et al., 2018 showed that given the headway constraints, the number of HDVs that a single autonomous vehicle can stabilize is given by (Equation 36):

$$\begin{aligned}n^*_{s/e} &= \min_{\omega}\{\min_n\{\left(log|1 - T_A(j\omega)| + \sum_i^n log|T_i(j\omega)|\right.\\ &\left. - log(\eta)\right) \cdot log|1 - T_A(j\omega)| + \sum_i^{n+1} log|T_i(j\omega)|\\ &- log(\eta) \leq 0\}\}\end{aligned} \quad (36)$$

where:

$\eta = \Delta/\beta$, represents the relative scale of the disturbance.

## 4   OPTIMIZATION PROBLEM FORMULATION

The overall controller design problem for CAVs can be formulated as a multi-objective problem with the following objectives: (a) maximize the number of HDVs that the CAV can stabilize given oscillations which may trigger stop-and-go waves, (b) minimize the risks of collision. The underlying settings of the problem can be summarized as follows:

- The problem considers that there exist certain conditions under which the HDVs are not string stable. This means that small perturbations from a uniform flow are amplified as they propagate from the leading vehicle to vehicles that follow it.

- Human-driven vehicle models are considered to be characterized by heterogeneity and time delay. The latter, caused by the length of human perception-reaction time, is of interest. The parameters that characterize human behaviors are not necessarily the same across individual human drivers.

- All the HDVs that are to be stabilized by the CAV are well-connected through electronic connectivity. In this paper, this assumption can be considered appropriate because in the study, we define this based on a specific range of vehicle-to-vehicle communication.

The optimization problem can be expressed as follows (Equations (37)-(40)):

$$max\ n^*_{stable}, n^*_{s/e} \quad (37)$$

s.t.

$$T_A(s) = \frac{(k_1 + sk_3)}{s^2 + s(k_2 + k_3 + k_1\lambda_2) + k_1} \quad (38)$$

$$k_2^2 + k_1^2\lambda_2^2 + 2k_2k_3 + 2k_1k_2\lambda_2 + 2k_1k_3\lambda_2 - 2k_1 \geq 0 \quad (39)$$

$$k_1, k_2, k_3 \geq 0 \quad (40)$$

where $k_1, k_2, k_3$ are decision variables to be tuned.

## 5 NUMERICAL EXPERIMENT

In the numerical experiment, we test initially using the following scenarios: the optimal velocity $v^* = 30$ mph. A specific optimal velocity function, (Equation 41) given by Bando (Bando et al., 1995), based on the car-following experiment by Koshi et al.

$$V(\Delta x) = 16.8[\tanh 0.0860(\Delta x - (20 + l_c)) + 0.913] \quad (41)$$

The length of the vehicle $l_c = 5m$, which was used in the Chu Motorway car-following experiment (Bando et al., 1995). Then the $V(\Delta x^*)'$ can be expressed as (Equation (42)):

$$V(\Delta x^*)' = 1.4448[1 - \tanh^2 0.0860(\Delta x^* - 25)] \quad (42)$$

Based on the California Vehicle Code, the constant time headway can be established as: $\lambda_2 = 0.225 l_c v_i = 1.125 v_i$.

Considering heterogeneity in human drivers behavior, the desired headways can be viewed as random variable across the population. Assume $v_0 = 45$ mph, $\Delta x_i^* = 1.5 v_i + 2 \sim \mathcal{N}(30.125, 2)$. Additionally, using some calibrated values for trade-off parameters $\alpha_i$ and $\beta_i$ given in (Jiang et al., 2001), $\alpha_i = 0.04 s^{-1}$, $\beta_i = 0.18 s^{-1}$. Based on these assumptions, we can define the human car-following model as shown in Equation (10), with the following parameters: $k_{1i} = \alpha_i V'(\Delta x_i^*)$, $k_{2i} = \alpha_i$, and $k_{3i} = \beta_i$. Also, we sample the relevant parameters for the HDV models from the following distributions (Equations (43)-(45)):

$$\alpha_i = 0.41 + \epsilon_\alpha \sim \mathcal{N}(0.04, 0.004) \quad (43)$$

$$\beta_i = 0.5 + \epsilon_\beta \sim \mathcal{N}(0.185, 0.018) \quad (44)$$

$$\Delta x_i^* \sim \mathcal{N}(30.125, 3) \quad (45)$$

The sensitivity parameter, $\alpha$, is such that $\alpha \propto 1/\tau_i$. Therefore, the time-delay for each HDV model is obtained by $\tau_i = 1/2500 \alpha_i$.

By tuning the parameters that characterize the controller, $k_1$, $k_2$ and $k_3$, the maximum number of HDVs that can be safely stabilized, is optimized at $k_1 = 0$, $k_2 \approx \eta k_3$ (as shown in Fig. 2).

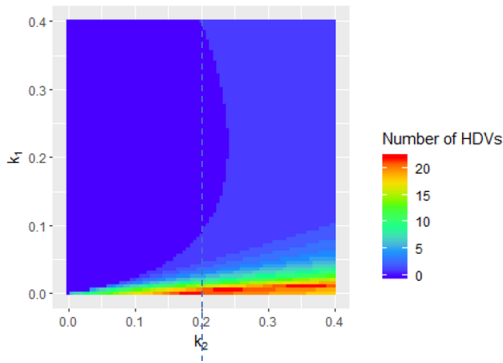

(a) $k_1$ and $k_2$

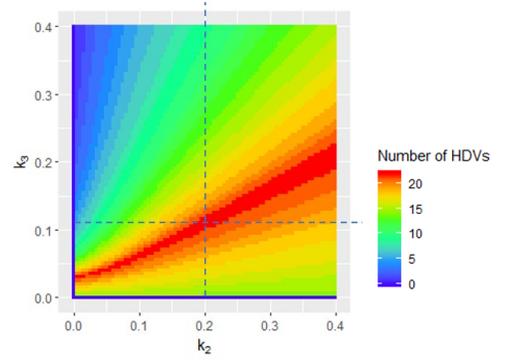

(b) $k_1$ and $k_3$

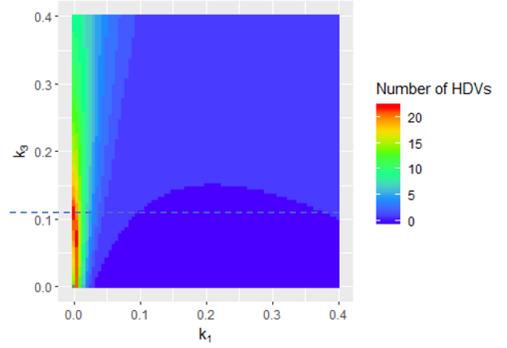

(c) $k_2$ and $k_2$

Fig. 2. Heatmaps of relationships between controller gains parameter combinations and the maximum number of HDVs that can be stabilized

Given the optimal controller for CAVs, we try to explore the impacts of frequency $\omega$ by find the trend of number of HDVs that can be stabilized with varied frequency. From the analysis, the most unfavorable frequency range can be determined (Fig. 3). For purposes of comparison, we also tested scenarios without considering time delay and heterogeneity in HDVs. (also shown in Fig. 3.). We found that the number of HDVs that a CAV can optimize is lower when human drivers potential time delay and heterogeneity are considered, compared to the scenario where such conditions (time delay and heterogeneity) are not considered. This result suggests that heterogeneity and time delay in HDV behavior impair the CAVs capability to stabilize traffic. Therefore, in designing CAV controllers for traffic stabilization, it is essential to consider such uncertainty-related conditions by adopting more realistic models of human driver behavior.

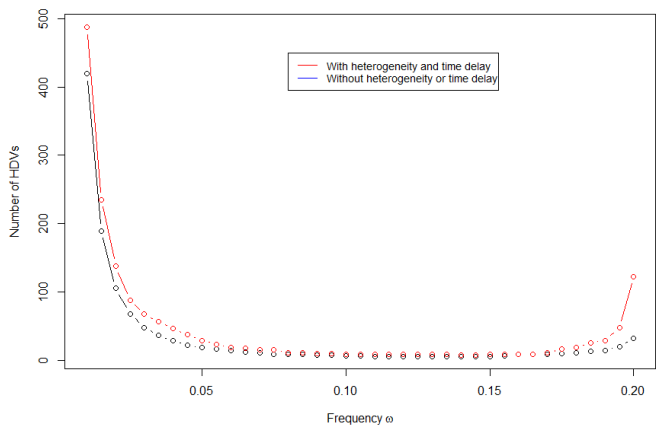

Fig. 3. Relationship between the maximum number of HDVs that can be stabilized and the disturbance frequency

## 6 DISCUSSION AND CONCLUSIONS

In this paper, we focused on the mixed traffic flow conditions, which consists of HDVs and CAVs. We modeled the HDVs with due consideration of driver heterogeneity and time delays. We designed the CAV controllers based on the output metrics of string stability, efficiency and safety. We demonstrated the impacts of time delay and heterogeneity behavior on the traffic performance outputs using mathematical proof and a numerical experiment. The analysis was repeated for homogeneous HDV platoons with zero response delay associated with perception reaction. The results indicate that due consideration of human behavior, and the variation of its sensitivity and driving patterns, yield results that are different compared to non-consideration of such behavior. The results also show that the maximum number of HDVs that a CAV can safely and effectively stabilize is lower when human drivers potential time delay and heterogeneity are considered, compared to the unrealistic scenario where such time delay and heterogeneity are not considered. Therefore, in a realistic driving environment, heterogeneity and time delay in HDV behavior impair the CAVs capability to stabilize traffic, and need to be considered in efforts that measure CAV traffic-stablizing efficacy.

One of the strengths of our proposed framework is that if the HDVs are connected, the dynamics of HDV models can be calibrated in real time. This provides the CAV controller the flexibility to capture heterogeneity in human driver behavior and to incorporate the effect of such flexibility. It also helps the CAV to predict the trajectory of upstream HDVs more precisely and in real time. Additionally, the framework is capable of modeling potential interactions between HDVs and CAVs. Since human drivers may have different level of acceptance of CAVs, their car-following behavior may differ from each other. Specifically, drivers who have lower trust in autonomous mobility may tend to adopt a conservative headways when they find the preceding vehicle is autonomous vehicle. In the numerical experiment part of our paper, we tested the proposed methodology with sampled data based on specific assumption regarding the distributions of the parameters. To test the robustness of the controller design and to make sure that the controller can handle the real-world complexity, it would be helpful to calibrate the parameters characterizing HDVs with empirical car-following data, e.g., the Next Generation SIMulation (NGSIM) data. Fig. 3 shows, a stop-and-go wave was triggered and propagated to the following (or, upstream) vehicles that travels backwards along the platoon. Further research can investigate calibration of HDVs models in real-time and seek to optimize the estimation error and time.

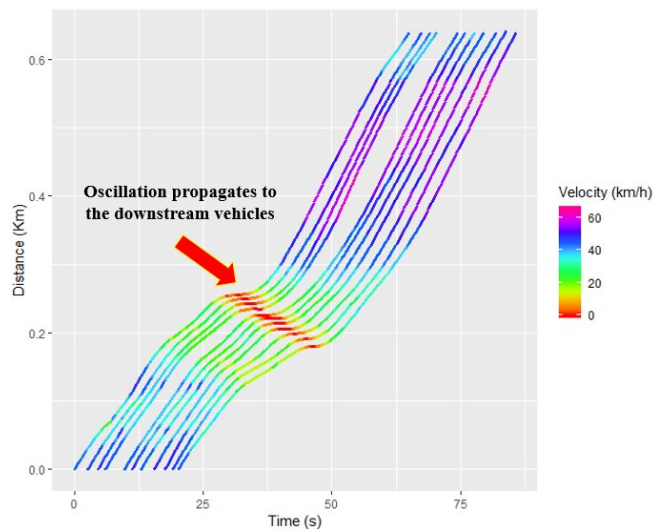

Fig. 4. Stop-and-go waves in the empirical data

This study has some limitations that should be mentioned here, to serve as a beacon not only for practitioners who wish to adopt the study results for implementation but also for future reseachers who wish to replicate the study and subsequently to improve the model further. In the paper, we assumed there is no communication time delay for the CAV models. However, in reality, there could exist a little delay in the information processing by CACC controllers. For example, due to the complexity of HDV model calibration, the computation process may be time-consuming and could introduce additional time delay into the controller system. Therefore, future research could assess the impact of such time delay on CACC controller performance, and subsequently evaluate the extent to which such delay could affect CAV efficacy in traffic string stabilization.

## ACKNOWLEDGEMENTS AND DISCLAIMER

This work was supported by Purdue University's Center for Connected and Automated Transportation (CCAT), a part of the larger CCAT consortium, a USDOT Region 5 University Transportation Center funded by the U.S. Department of Transportation, Award #69A3551747105. The contents of this paper reflect the views of the authors, who are responsible for the facts and the accuracy of the data presented herein, and do not necessarily reflect the official views or policies of the sponsoring organization.

## REFERENCES

Bando, M., Hasebe, K., Nakanishi, K., Nakayama, A., 1998. Analysis of optimal velocity model with explicit delay. Phys. Rev. E - Stat. Physics, Plasmas, Fluids, Relat. Interdiscip. Top. 58, 5429–5435. https://doi.org/10.1103/PhysRevE.58.5429

Bando, M., Hasebe, K., Nakanishi, K., Nakayama, A., Shibata, A., Sugiyama, Y., 1995. Phenomenological study of dynamical model of traffic flow. J. Phys. I 5, 1389–1399.

Bando, M., Hasebe, K., Nakayama, A., Shibata, A., Sugiyama, Y., 1994. Structure stability of congestion in traffic dynamics. Jpn. J. Ind. Appl. Math. 11(2), 203-210.

Bekey, G.A., Burnham, G.O., Seo, J., 1977. Control Theoretic Models of Human Drivers in Car Following. Hum. Factors J. 19(4), 399-413.

Burnham, G.O., Seo, J., Bekey, G.A., 1974. Identification of Human Driver Models in Car Following. IEEE Trans. Automat. Contr. 19(6), 911-915.

Chandler, R.E., Herman, R., Montroll, E.W., 1958. Traffic Dynamics: Studies in Car Following. Oper. Res. 6(2), 165-184.

Chen, S., Li, Y., Du, R., Dong, J., Ha, Y.J. Labi, S. (2020). Collision avoidance framework for autonomous vehicles under crash imminent situations. National Research-review webinsr series, Center for Connected and Automatd Transporattion, University of Michigan Transportation Research Institute.

Chen, S., Leng, Y., Labi, S., 2020. A deep learning algorithm for simulating autonomous driving considering prior knowledge and temporal information. Comput. Civ. Infrastruct. Eng. https://doi.org/10.1111/mice.12495

Dong, J., Chen, S., Li, Y., Du, R., Steinfeld, A., Labi, S. (2020). Spatio-weighted information fusion and DRL-based control for connected autonomous vehicles, IEEE ITS Conference, September 20–23, 2020. Rhodes, Greece.

Gunter, G., Janssen, C., Barbour, W., Stern, R.E., Work, D.B., 2019. Model-Based String Stability of Adaptive Cruise Control Systems Using Field Data. IEEE Trans. Intell. Veh. https://doi.org/10.1109/tiv.2019.2955368

Hedrick, J.K., McMahon, D., Narendran, V., Swaroop, D., 1991. Longitudinal vehicle controller design for IVHS system, in: Proceedings of the American Control Conference, 3107-3112, https://doi.org/10.23919/acc.1991.4791980

Horn, B.K.P., Wang, L., 2018. Wave Equation of Suppressed Traffic Flow Instabilities. IEEE Trans. Intell. Transp. Syst. https://doi.org/10.1109/TITS.2017.2767595

Hydén, C., 1996. Traffic conflicts technique: state-of-the-art. Traffic Saf. Work with video Process. 37, 3–14.

Ioannou, P., Xu, Z., 1994. Throttle and brake control systems for automatic vehicle following. IVHS J. 1, 345–377.

Ioannou, P.A., Chien, C.C., 1993. Autonomous intelligent cruise control. IEEE Trans. Veh. Technol. 42, 657–672. https://doi.org/10.4271/930510

Jiang, R., Hu, M. Bin, Zhang, H.M., Gao, Z.Y., Jia, B., Wu, Q.S., Wang, B., Yang, M., 2014. Traffic experiment reveals the nature of car-following. PLoS One 9(4). https://doi.org/10.1371/journal.pone.0094351

Jiang, R., Jin, C.J., Zhang, H.M., Huang, Y.X., Tian, J.F., Wang, W., Hu, M. Bin, Wang, H., Jia, B., 2018. Experimental and empirical investigations of traffic flow instability. Transp. Res. Part C Emerg. Technol. 94(1), 83-98.

Jiang, R., Wu, Q., Zhu, Z., 2001. Full velocity difference model for a car-following theory. Phys. Rev. E, 64(1), 17101.

Kato, S., Tsugawa, S., Tokuda, K., Matsui, T., Fujii, H., 2002. Vehicle Control Algorithms for Cooperative Driving with Automated Vehicles and Intervehicle Communications. IEEE Trans. Intell. Transp. Syst. , 155-161. https://doi.org/10.1109/TITS.2002.802929

Krishnan, H., Gibb, S., Steinfeld, A., Shladover, S., 2001. Rear-end collision-warning system: Design and evaluation via simulation, in: Transportation Research Record. 1759(1), 52-60.

Laval, J.A., 2006. Stochastic processes of moving bottlenecks: Approximate formulas for highway capacity. Transp. Res. Rec. 1988, 86–91.

Laval, J.A., Daganzo, C.F., 2006. Lane-changing in traffic streams. Transp. Res. Part B Methodol. 40(3), 251–264.

Litman, T., 2019. Autonomous Vehicle Implementation Predictions: Implications for Transport Planning. Transp. Res. Board Annu. Meet. https://doi.org/10.1613/jair.301

Milanes, V., Shladover, S.E., Spring, J., Nowakowski, C., Kawazoe, H., Nakamura, M., 2014. Cooperative adaptive cruise control in real traffic situations. IEEE Trans. Intell. Transp. Syst. 15, 296–305. https://doi.org/10.1109/TITS.2013.2278494

Milanés, V., Shladover, S.E., Spring, J., Nowakowski, C., Kawazoe, H., Nakamura, M., 2013. Cooperative adaptive cruise control in real traffic situations. IEEE Trans. Intell. Transp. Syst. 15, 296–305.

Minderhoud, M.M., Bovy, P.H.L., 2001. Extended time-to-collision measures for road traffic safety assessment. Accid. Anal. Prev., 33(1), 89-97. https://doi.org/10.1016/S0001-4575(00)00019-1

Nakayama, A., Sugiyama, Y., Hasebe, K., 2002. Effect of looking at the car that follows in an optimal velocity model of traffic flow. Physical Review E, 65(1), 016112. https://doi.org/10.1103/PhysRevE.65.016112

Orosz, G., Wilson, R.E., Stépán, G., 2010. Traffic jams: dynamics and control.

Orosz, G., Wilson, R.E., Szalai, R., Stépán, G., 2009. Exciting traffic jams: Nonlinear phenomena behind traffic jam formation on highways. Phys. Rev. E 80, 46205.

Pipes, L.A., 1953. An operational analysis of traffic dynamics. J. Appl. Phys., 24(3), 274-281. https://doi.org/10.1063/1.1721265

Schakel, W.J., Van Arem, B., Netten, B.D., 2010. Effects of cooperative adaptive cruise control on traffic flow stability, Proceedings IEEE Conference on Intelligent Transportation Systems, 759-764, ITSC. https://doi.org/10.1109/ITSC.2010.5625133

Schrank, D., Lomax, T., & Eisele, B. (2012). 2012 urban mobility report. Texas Transportation Institute,

http://mobility.tamu.edu/ums/report.
Sheikholeslam, S., Desoer, C.A., 1992. Control of Interconnected Nonlinear Dynamical Systems: The Platoon Problem. IEEE Trans. Automat. Contr., 37(6), 806-810. https://doi.org/10.1109/9.256337
Shladover, S., VanderWerf, J., Miller, M.A., Kourjanskaia, N., Krishnan, H., 2001. Development and performance evaluation of AVCSS deployment sequences to advance from today's driving environment to full automation, Report Nr. UCB-ITS-PRR-2001-18, California Partners For Advanced Transit And Highways (PATH) Research Institute Of Transportation Studies, University Of California, Berkeley, CA.
Shladover, S.E., Su, D., Lu, X.Y., 2012. Impacts of cooperative adaptive cruise control on freeway traffic flow. Transp. Res. Rec. 2324(1), 63-70. https://doi.org/10.3141/2324-08
Stern, R.E., Cui, S., Delle Monache, M.L., Bhadani, R., Bunting, M., Churchill, M., Hamilton, N., Haulcy, R., Pohlmann, H., Wu, F., Piccoli, B., Seibold, B., Sprinkle, J., Work, D.B., 2018. Dissipation of stop-and-go waves via control of autonomous vehicles: Field experiments. Transp. Res. Part C Emerg. Technol. 89, 205-221. https://doi.org/10.1016/j.trc.2018.02.005
Sugiyamal, Y., Fukui, M., Kikuchi, M., Hasebe, K., Nakayama, A., Nishinari, K., Tadaki, S.I., Yukawa, S., 2008. Traffic jams without bottlenecks-experimental evidence for the physical mechanism of the formation of a jam. New J. Phys., 10(3), 033001. https://doi.org/10.1088/1367-2630/10/3/033001
Swaroop, D., 1996. String Stability of Interconnected Systems - Automatic Control, IEEE Transactions on. IEEE Trans. Automat. Contr. 41, 349–357.
Tadaki, S.I., Kikuchi, M., Fukui, M., Nakayama, A., Nishinari, K., Shibata, A., Sugiyama, Y., Yosida, T., Yukawa, S., 2013. Phase transition in traffic jam experiment on a circuit. New J. Phys. 15(10), 103034. https://doi.org/10.1088/1367-2630/15/10/103034
Talebpour, A., Mahmassani, H.S., 2016. Influence of connected and autonomous vehicles on traffic flow stability and throughput. Transp. Res. Part C Emerg. Technol. 71, 143–163.
Tarko, A.P., 2018. Estimating the expected number of crashes with traffic conflicts and the Lomax Distribution – A theoretical and numerical exploration. Accid. Anal. Prev. 13, 63-73. https://doi.org/10.1016/j.aap.2018.01.008
Tyler, J.S., 1964. The Characteristics of Model-Following Systems as Synthesized by Optimal Control. IEEE Trans. Automat. Contr. 9(4), 485-498. https://doi.org/10.1109/TAC.1964.1105757
US Energy Information Administration, 2017. Study of the potential energy consumption impacts of connected and automated vehicles. Washington, DC.
Vahidi, A., & Eskandarian, A. (2003). Research advances in intelligent collision avoidance and adaptive cruise control. IEEE transactions on intelligent transportation systems, 4(3), 143-153.
van den Broek, T.H.A., Ploeg, J., Netten, B.D., 2011. Advisory and Autonomous Cooperative Driving Systems. IEEE Int. Conf. Consum. Electron. 279–280.
VanderWerf, J., Shladover, S.E., Miller, M.A., Kourjanskaia, N., 2002. Effects of adaptive cruise control systems on highway traffic flow capacity. Transp. Res. Rec. 1800(1), 78-84. https://doi.org/10.3141/1800-10
Wilmink, I.R., Klunder, G.A., Van Arem, B., 2007. Traffic flow effects of Integrated full-Range Speed Assistance (IRSA), in: IEEE Intelligent Vehicles Symposium, Proceedings, 1204-1210. https://doi.org/10.1109/ivs.2007.4290282
Wilson, R.E., Ward, J.A., 2011. Car-following models: fifty years of linear stability analysis--a mathematical perspective. Transp. Plan. Technol. 34, 3–18.
Wu, C., Bayen, A.M., Mehta, A., 2018. Stabilizing Traffic with Autonomous Vehicles, in: Proceedings - IEEE International Conference on Robotics and Automation. https://doi.org/10.1109/ICRA.2018.8460567
Zheng, L., Ismail, K., Meng, X., 2014. Traffic conflict techniques for road safety analysis: open questions and some insights. Can. J. Civ. Eng. 41, 633–641.
Zheng, Z., Ahn, S., Chen, D., Laval, J., 2011. Applications of wavelet transform for analysis of freeway traffic: Bottlenecks, transient traffic, and traffic oscillations. Transp. Res. Part B Methodol. 45, 372–384.